\documentclass{article}
\usepackage[utf8]{inputenc}
\usepackage{libertine}
\usepackage{amsthm,amsmath,amssymb,amsfonts}
\usepackage{bbm} 

\def\Real{\mathbb{R}}
\def\sigmaB{\mathcal{B}}
\def\sigmaF{\mathcal{F}}
\def\sigmaS{\mathcal{S}}
\def\Prob{\mathbf{P}}
\def\ind{\mathbbm{1}}
\newcommand\Expect[1]{\mathbf{E} \left[ #1 \right]}
\def\mix{\mathop\mathrm{mix}\limits}
\DeclareMathOperator{\ES}{ES}
\def\Set#1#2{\left\{{#1}\ \colon\ {#2} \right\}}

\DeclareMathOperator*{\essinf}{ess\,inf}
\newcommand{\defeq}{\mathrel{\mathop:}=}

\newtheoremstyle{my_def_style} 
{5pt} 
{3pt} 
{} 
{} 
{\scshape\bfseries} 
{.} 
{.5em} 
{} 

\newtheoremstyle{my_thm_style} 
{5pt} 
{3pt} 
{\itshape} 
{} 
{\scshape\bfseries} 
{.} 
{.5em} 
{} 

\theoremstyle{my_def_style}
\theoremstyle{my_thm_style}
\newtheorem*{my_lemma}{Lemma}
\newtheorem*{my_thm}{Theorem}

\newenvironment{my_prf}[1][\proofname]{\par\noindent\pushQED{\qed}\scshape\bfseries#1. \normalfont\ignorespaces}{\popQED}

\title{On the Concavity of Expected Shortfall}
\author{Mikhail Tselishchev \thanks{Department of Mathematical Statistics, Faculty of Computational Mathematics and Cybernetics, Moscow State University. Email: \textit{mihail.tselishchev(at)gmail(dot)com}.}}
\date{}

\begin{document}

\maketitle
\thispagestyle{empty}  

\begin{abstract}
It is well known that Expected Shortfall (also called Average Value-at-Risk) is a convex risk measure, 
i.\,e. Expected Shortfall of a convex linear combination of arbitrary risk positions is not greater than
a convex linear combination with the same weights of Expected Shortfalls of the same risk positions.
In this short paper we prove that Expected Shortfall is a~concave risk measure 
with respect to probability distributions, 
i.\,e. Expected Shortfall of~a~finite mixture of arbitrary risk positions is not lower than 
the linear combination of Expected Shortfalls of the same risk positions (with the same weights as in the mixture).
\end{abstract}

\section{Introduction}
Expected Shortfall ($\ES$) is a standard risk measure used by financial institutions and regulators to determine capital requirements. 
The notion of Expected Shortfall was introduced in \cite{AcerbiTasche_ES,AcerbiTasche_OnTheCoherence}.
Expected Shortfall at level~$\alpha$ is defined as an average loss in the worst $\alpha \cdot 100 \%$ cases for a given risk position. 

In order to give mathematical definitions we will need some basic notation. 
We assume that risk positions are random variables on some probability space $(\Omega,\sigmaF,\Prob)$, they denote profits (or losses when negative) of some asset or portfolio at the end of the trading period.
We will denote by $F_X(\cdot)$ the cumulative distribution function (cdf) of risk position $X$. 
A lower quantile function (also known as an inverse cdf) for the distribution of $X$ is defined as
\begin{equation}
q_{\alpha}(X) \equiv F_X^{-}(\alpha) \defeq \inf \Set{x \in \overline\Real}{F_X(x) \ge \alpha}, \quad \alpha\in[0,1].
\end{equation}
Note that this definition is independent of whether one takes left-continuous or right-continuous cdf's.
In the following definition of $\ES$, one can replace lower quantiles with the upper quantiles
\begin{equation}
q^{(\alpha)}(X)  \defeq \inf \Set{x \in \overline\Real}{F_X(x) > \alpha}, \quad \alpha\in[0,1),
\end{equation}
since $\ES$ is an integral characteristic of quantile function, and the~lower quantile function coincides with the~upper quantile function almost everywhere w.r.t. Lebesgue measure on~$[0,1]$. 
More precisely, they differ only on at most countable set of points $\alpha\in[0,1]$, where $F_X^{-1}(\{\alpha\})$ consists of at least two different points (and thus the~whole interval of~constancy of~$F_X$). One might check that the lower quantile function is left-continuous, while the upper quantile function is right-continuous.

Expected Shortfall of risk position $X$ at level $\alpha\in(0,1]$ is defined as
\begin{equation}\label{eq:ES:def1}
\ES_\alpha(X) \defeq - \frac{1}{\alpha} \int_0^\alpha q_\gamma(X) \, d\gamma,
\end{equation}
which makes clear why $\ES$ is sometimes called as Average Value-at-Risk (AVaR).
It is well known that $q_\gamma(X)$, as a function of $\gamma$ on the unit interval with Lebesgue measure, has the same probability distribution as risk position $X$ itself. This remark clarifies the fact that $\ES_\alpha(X)$ shows the average loss in the worst $\alpha \cdot 100 \%$ cases, since ``cases'' are naturally ordered on the unit interval, and the worst $\alpha \cdot 100 \%$ cases for $q_\gamma(X)$ are located on the~interval $(0,\alpha)$.

Alternatively, Expected Shortfall can be defined a bit trickier:
\begin{equation}\label{eq:ES:def2}
\ES_\alpha(X) \defeq -\frac{1}{\alpha} \biggl( \Expect{X\ind_{\left\{X < q_\alpha(X)\right\}}}
+ q_\alpha(X) \cdot \bigl( \alpha - \Prob{(X < q_\alpha(X) } \, \bigr) 
\  \biggr).
\end{equation}
We will use both representations~\eqref{eq:ES:def1} and~\eqref{eq:ES:def2} in our work.
The proof of their equivalence can be found in~\cite{AcerbiTasche_OnTheCoherence}.

Note that $\ES_\alpha(X)$ is correctly defined for risk positions $X$ with $\Expect{X^-} < +\infty$.
We will assume that this condition is satisfied for all risk positions presented in~the~paper.

According to~\eqref{eq:ES:def1} (or~\eqref{eq:ES:def2}), $\ES_1(X)$ coincides with $-\Expect{X}$.
Expected Shortfall at level~$0$ is often defined as
\begin{equation}\label{eq:ES:level0}
\ES_0(X) \defeq - \essinf X.
\end{equation}

Here and later we will use the notation of a standard $(n{-}1)$-simplex: 
\begin{equation*}
\sigmaS_{n-1} \defeq \Set{(x_1,\ldots,x_n) \in \Real^n}{x_i \ge 0 \text{ for all } i=1,\ldots, n \text{, and } \sum_{i=1}^n x_i = 1}.
\end{equation*}

Expected Shortfall is known to be convex, i.e. for  random vector $X = (X_1,\ldots,X_n)$, weights $\beta\in\sigmaS_{n-1}$ and arbitrary level $\alpha\in[0,1]$, the following inequality holds true:
\begin{equation}\label{eq:ES:convexity}
\ES_\alpha\left(\sum_{j=1}^n \beta_j X_j \right)  \le \sum_{j=1}^n \beta_j \ES_\alpha(X_j),
\end{equation}
which states that the risk of a convex linear combination of risk positions 
cannot exceed the convex linear combination (with the same weights) of marginal risks.
Due to positively homogeneity of Expected Shortfall (that is $\ES_\alpha (\lambda X) = \lambda\ES_\alpha (X)$ for any $\lambda\ge 0$, this fact can be proved directly),  inequality~\eqref{eq:ES:convexity} is equivalent to the subadditivity of Expected Shortfall:
$\ES_\alpha(X+Y) \le \ES_\alpha(X) + \ES_\alpha(Y)$. One can find several proofs of subadditivity in~\cite{EmbrechtsWang_SevenProofs}. 

The equality in~\eqref{eq:ES:convexity} is attained when ``worst cases'' for all risk positions coincide and have the same ordering up to level $\alpha$. In particular, this is the case when all~$X_j$ are \textit{comonotone}, i.\,e. their joint copula has the form of $\min(x_1,\ldots,x_n)$ for $x_1,\ldots,x_n \in [0,1]$\ .

We will denote a mixture of random vector $X = (X_1,\ldots,X_n)$ with weights $\beta\in \sigmaS_{n-1}$ as $\mix_\beta X$. This random variable, whose cdf is $\sum_{j=1}^n \beta_j F_{X_j}(\cdot)$, is not uniquely defined (or even could not be defined at all on $(\Omega,\sigmaF,\Prob)$ if this probability space is too poor), but since Expected Shortfall and other quantile-based risk measures are characteristics of distribution of random variable, there should be no confusion in what follows.

\section{Main Result}

We will use the following lemma to prove the concavity of Expected Shortfall.

\begin{my_lemma}
Let $X=(X_1,\ldots,X_n)$ be a random vector and $\beta \in \sigmaS_{n-1}$. Then for every $\alpha \in (0,1)$ there exist $\alpha_1,\ldots,\alpha_n \in [0,1]$, such that $\alpha = \sum_{j=1}^n \alpha_j \beta_j$ and 
\begin{equation} \label{eq:lemma}
\ES_\alpha (\mix_\beta X) = \sum_{j=1}^n \frac{\alpha_j \beta_j}{\alpha} \ \ES_{\alpha_j}(X_j).
\end{equation}
\end{my_lemma}
\begin{my_prf}
Let $\xi = \mix_\beta X$, and $q_\alpha=q_\alpha(\xi)$ be an $\alpha$-lower quantile of this mixture. 
Without loss of generality we assume all $\beta_j > 0$, $j = 1 \ldots n$.
We are going to show that the following $\alpha_j$ satisfy the statement of lemma:
\begin{equation} \label{eq:lemma:alpha_j}
\alpha_j \defeq \Prob(X_j < q_\alpha) + \Prob(X_j = q_\alpha) \cdot \frac{\alpha - \Prob(\xi < q_\alpha)}{\Prob(\xi = q_\alpha)}.
\end{equation}
Note that if $\Prob(\xi = q_\alpha) = 0$, then the cdf of $\xi$ is continuous at $q_\alpha$, and thus the cdf of each $X_j$ is also continuous at that point, i.\,e. $\Prob(X_j = q_\alpha) = 0$, and the second term in the right-hand side of~\eqref{eq:lemma:alpha_j} should be interpreted as zero.
From the definition of the lower quantile function, one has $\alpha_j\in \left[\Prob(X_j < q_\alpha), \Prob(X_j \le q_\alpha)\right]$. Next, if $\Prob(\xi = q_\alpha) > 0$, then
\begin{equation*}
\sum_{j=1}^n \alpha_j \beta_j = \Prob(\xi < q_\alpha) + \Prob(\xi = q_\alpha) \cdot \frac{\alpha - \Prob(\xi < q_\alpha)}{\Prob(\xi = q_\alpha)} = \alpha.
\end{equation*}
If $\Prob(\xi = q_\alpha) = 0$, then $\Prob(\xi < q_\alpha) = \alpha$, and $\sum_{j=1}^n \alpha_j \beta_j = \Prob(\xi < q_\alpha) = \alpha$ as~well.
It only remains to prove~\eqref{eq:lemma}. 
According to~\eqref{eq:ES:def2}, the latter equality and properties of the mixture of distributions, one has
\begin{equation} \label{eq:lemma:add1}
\begin{split}
\ES_\alpha(\xi) &= - \frac{1}{\alpha} \left( \bigg. \Expect{\xi \ind_{\{\xi < q_\alpha\}}} + q_\alpha \cdot \left(\Big. \alpha - \Prob\left(\xi < q_\alpha\right) \right) \right)
=\\
&= - \frac{1}{\alpha}\left( \sum_{j=1}^n \beta_j \, \Expect{X_j \ind_{\{X_j < q_\alpha\}}} + \sum_{j=1}^n \beta_j \, q_\alpha \cdot \left( \Big. \alpha_j - \Prob\left(X_j < q_\alpha\right) \right)\right)
=\\
&= - \frac{1}{\alpha} \sum_{j=1}^n \beta_j \left( \Bigg. \Expect{X_j \ind_{\{X_j < q_\alpha\}}} +  q_\alpha \cdot \left( \Big. \alpha_j - \Prob\left(X_j < q_\alpha \right) \right)\right).
\end{split}
\end{equation}
If some $\alpha_j=0$, then $\Prob(X_j < q_\alpha) = 0$, which implies $\ES_0(X_j) < +\infty$ and
\begin{equation} \label{eq:lemma:add1_1}
\Expect{X_j \ind_{\{X_j < q_\alpha\}}} +  q_\alpha \cdot \left( \Big. \alpha_j - \Prob\left(X_j < q_\alpha \right) \right) = 0 = -\alpha_j \ES_{\alpha_j}(X_j)
\end{equation}
We want to show that for every $j$ with $\alpha_j > 0$ the following equality holds true:
\begin{equation} \label{eq:lemma:add2}
\begin{split}
 & \Expect{X_j \ind_{\{X_j < q_\alpha\}}} +  q_\alpha \cdot \left( \Big. \alpha_j - \Prob\left(X_j < q_\alpha \right) \right) =
\\= \
& \Expect{X_j \ind_{\{X_j < q_{\alpha_j}(X_j)\}}} +  q_{\alpha_j}(X_j) \cdot \left( \Big. \alpha_j - \Prob\left(X_j < q_{\alpha_j}(X_j) \right) \right).
\end{split}
\end{equation}
Since $\Prob(X_j \le q_\alpha) \ge \alpha_j$, then $q_{\alpha_j}(X_j) \le q_\alpha$. 
If $q_{\alpha_j}(X_j) = q_\alpha$, then~\eqref{eq:lemma:add2} is~trivial.
If, however, $q_{\alpha_j}(X_j) < q_\alpha$, then one has a chain of inequalities:
\begin{equation*}
\Prob(X_j < q_\alpha) \ge \Prob(X_j \le q_{\alpha_j}(X_j)) \ge \alpha_j \ge \Prob(X_j < q_\alpha).
\end{equation*}
Hence, $\Prob(X_j \le q_{\alpha_j}(X_j)) = \alpha_j$ and $\Prob( q_{\alpha_j}(X_j) < X_j < q_\alpha) = 0$.
Given that, consider the difference between the left- and the right-hand sides of~\eqref{eq:lemma:add2}:
\begin{equation*}
\begin{split}
 & \Expect{X_j \ind_{\{X_j < q_\alpha\}}} +  q_\alpha \cdot \left( \Big. \alpha_j - \Prob\left(X_j < q_\alpha \right) \right) -
\\& - \
 \Expect{X_j \ind_{\{X_j < q_{\alpha_j}(X_j)\}}} -  q_{\alpha_j}(X_j) \cdot \left( \Big. \alpha_j - \Prob\left(X_j < q_{\alpha_j}(X_j) \right) \right)=
\\= \  
&\Expect{X_j \ind_{\{q_{\alpha_j}(X_j) \le X_j < q_\alpha\}}}
+ q_\alpha \cdot \left( \Big. \alpha_j - \Prob\left(X_j < q_\alpha \right) \right) - \\& - q_{\alpha_j}(X_j) \cdot \left( \Big. \alpha_j - \Prob\left(X_j < q_{\alpha_j}(X_j) \right) \right)=
\\= \ 
& q_{\alpha_j}(X_j) \cdot \left( \Big. \Prob(X_j \le q_{\alpha_j}(X_j)) - \alpha_j \right) + q_\alpha \cdot \left( \Big. \alpha_j - \Prob\left(X_j \le q_{\alpha_j}(X_j) \right) \right) =
\\= \ 
& \left( \Big. q_\alpha - q_{\alpha_j}(X_j) \right) \cdot \left( \Big. \alpha_j - \Prob\left(X_j \le q_{\alpha_j}(X_j) \right) \right) = 0.
\end{split}
\end{equation*}
Equation~\eqref{eq:lemma:add2} is proved.
Finally, \eqref{eq:lemma:add1}, together with~\eqref{eq:lemma:add1_1}, \eqref{eq:lemma:add2} and~\eqref{eq:ES:def2}, gives
\begin{equation*}
\begin{split}
\ES_\alpha(\xi) &= - \frac{1}{\alpha}  \sum_{j=1}^n \beta_j \left( \Bigg. \Expect{X_j \ind_{\{X_j < q_{\alpha_j}(X_j)\}}} +  q_{\alpha_j}(X_j) \left( \Big. \alpha_j - \Prob\left(X_j < q_{\alpha_j}(X_j) \right) \right)\right)
\\
&= \sum_{j=1}^n \frac{\beta_j \alpha_j}{\alpha} \ \ES_{\alpha_j} (X_j).
\end{split}
\end{equation*}
Lemma is proved.
\end{my_prf}

\noindent Now we are ready to formulate and prove the main result of the paper.

\begin{my_thm}[Concavity of Expected Shortfall] Let $X=(X_1,\ldots,X_n)$ be a random vector and $\beta \in \sigmaS_{n-1}$. Then for every $\alpha \in [0,1]$ one has
\begin{equation} \label{eq:ES:concavity}
\ES_\alpha \left( \mix_\beta X \right) \ge \sum_{j=1}^n \beta_j \ES_\alpha (X_j).
\end{equation}
\end{my_thm}
\begin{my_prf}
Again, we assume all $\beta_j > 0$ without loss of generality. If $\alpha = 0$, then
\begin{equation*}
\begin{split}
\ES_0(\mix_\beta X) &= -\essinf(\mix_\beta X) = - \min_{j=1\ldots n} \essinf(X_j)
\ge\\
&\ge - \sum_{j=1}^n \beta_j \essinf(X_j) = \sum_{j=1}^n \beta_j \ES_\alpha (X_j).
\end{split}
\end{equation*}
If $\alpha=1$, then
\begin{equation*}
\ES_1(\mix_\beta X) = -\Expect{\mix_\beta X} = -\sum_{j=1}^n \beta_j \Expect{X_j} = \sum_{j=1}^n \beta_j \ES_1(X_j).
\end{equation*}
In case $\alpha\in(0,1)$ we will take advantage of the previous lemma. Taking $q_\alpha$ and~$\alpha_j$ from its proof and using representation~\eqref{eq:ES:def1}, one has
\begin{multline} \label{eq:th:add1}
\ES_\alpha \left( \mix_\beta X \right)
= \sum_{j=1}^n \frac{\beta_j \alpha_j}{\alpha} \ \ES_{\alpha_j} (X_j)
= - \sum_{j=1}^n \frac{\beta_j}{\alpha} \ \int_0^{\alpha_j} q_\gamma(X_j) \, d \gamma =
\\ 
= \sum_{j=1}^n \beta_j \ES_\alpha(X_j) 
- \sum_{j\in J_+} \frac{\beta_j}{\alpha}\int_{\alpha}^{\alpha_j} q_\gamma(X_j) \, d \gamma
+ \sum_{j\in J_-} \frac{\beta_j}{\alpha}\int_{\alpha_j}^{\alpha} q_\gamma(X_j) \, d \gamma,
\end{multline}
where $J_+ = \Set{j}{\alpha_j > \alpha}$ and $J_- = \Set{j}{\alpha_j < \alpha}$.
For $j \in J_+$ we have
\begin{equation} \label{eq:th:add2}
\int_{\alpha}^{\alpha_j} q_\gamma(X_j) \, d \gamma 
\le q_{\alpha_j}(X_j) \cdot (\alpha_j - \alpha)
\le q_\alpha \cdot (\alpha_j - \alpha).
\end{equation}
Now consider $j\in J_-$. If $q_\gamma(X_j) < q_\alpha$ for some $\gamma > \alpha_j$, 
then $\Prob(X_j < q_\alpha) \ge \gamma > \alpha_j$, which leads to contradiction with the definition of~$\alpha_j$.
Thus, $q_\gamma(X_j) \ge q_\alpha$ for all $\gamma > \alpha_j$, and
\begin{equation} \label{eq:th:add3}
\int_{\alpha_j}^{\alpha} q_\gamma(X_j) \, d \gamma 
\ge q_\alpha \cdot (\alpha - \alpha_j).
\end{equation}
By substituting inequalities~\eqref{eq:th:add2} and~\eqref{eq:th:add3} into~\eqref{eq:th:add1}, we obtain
\begin{equation*}
\ES_\alpha \left( \mix_\beta X \right) 
\ge \sum_{j=1}^n \beta_j \ES_\alpha(X_j) 
+ \sum_{j=1}^n \frac{\beta_j}{\alpha} q_\alpha \cdot (\alpha - \alpha_j) 
= \sum_{j=1}^n \beta_j \ES_\alpha(X_j),
\end{equation*}
where the last equality holds due to $\beta\in\sigmaS_{n-1}$ and $\sum_{j=1}^n \beta_j \alpha_j = \alpha$.
\\The~theorem is proved.
\end{my_prf}

We have to notice that the proofs of both lemma and theorem are much simpler and more straightforward when all cdfs $F_{X_j}$ are continuous and strictly increasing.

Also note that inequality~\eqref{eq:ES:concavity} turns into equality in case all~$\alpha_j$ are equal to~$\alpha$. This means that $\alpha$-quantiles for all risk positions $X_1,\ldots,X_n $ coincide.

Proposed result can be naturally extended to a wider class of spectral risk measures.
Indeed, according to a spectral representation theorem (see~\cite{Acerbi_SpectralRepresentationTh} and~\cite{Kusuoka_OnLawInvariantCoherentRiskMeasures}), under suitable conditions of integrability, a~spectral risk measure may be represented as a~weighted combination of Expected Shortfalls
\begin{equation*}
\rho_\nu (X) = \int_{[0,1]} \ES_\alpha(X) \, d \nu(\alpha)
\end{equation*}
for some probability measure $\nu$ on $\left( [0,1], \sigmaB_{[0,1]} \right)$. 
Hence, using~\eqref{eq:ES:concavity},
\begin{equation*}
\begin{split}
\rho_\nu \left( \mix_\beta X \right)
&= \int_{[0,1]} \ES_\alpha \left( \mix_\beta X \right) \, d \nu(\alpha)
\ge \\ 
&\ge \sum_{j=1}^n \beta_j \int_{[0,1]} \ES_\alpha(X_j) \, d \nu(\alpha)
= \sum_{j=1}^n \beta_j \rho_\nu (X_j),
\end{split}
\end{equation*}
which means that the spectral risk measures are also concave with respect to~probability distributions.

Combining both~\eqref{eq:ES:convexity} and~\eqref{eq:ES:concavity}, one gets
\begin{equation} \label{eq:es_divers}
\ES_\alpha \left( \sum_{j=1}^n \beta_j X \right) \le \ES_\alpha \left( \mix_\beta X \right),
\end{equation}
i.\,e. a risk of a convex linear combination can not exceed a risk of a mixture of the same positions with the same weights. That useful property of $\ES$ (and spectral risk measures as well) might be called \textit{a principle of diversification}. This principle seems quite reasonable (unless one chooses a beverage in the bar).
When $\alpha$-quantiles of all risk positions coincide, and the ``worst cases'' up to level $\alpha$ are also the same for all of them, then inequality~\eqref{eq:es_divers} turns into equality.

\section{Conclusions}
In this paper we have shown a neat property of Expected Shortfall, which is concavity with respect to probability distributions. We have extended this property to a~class of~spectral risk measures. Finally, we have discussed the implications of concavity, that led us to the~principle of~diversification.
In our subsequent paper we are going to show that the principle of diversification is somewhat necessary for risk positions to be comparable with Expected Shortfall.

\section{Final words}
After finishing this paper, we found that the concavity of Expected Shortfall has already been proved recently in~\cite[Proposition 3.2]{Uryasev_FittingMixtureModels}. 
Their proof is even more elegant and compact. 
It uses another representation of Expected Shortfall, called Conditional Value-at-Risk (CVaR),  which is a solution to a specific minimization problem.
We decided to share our proof as well, since it differs internally. 
Moreover, our lemma might be useful in some other cases.

\bibliographystyle{ieeetr}
\bibliography{es_concavity}

\end{document}